\documentclass[a4paper,12pt]{article}
\usepackage{jheppub} 
\usepackage{ dsfont }
\usepackage{mathrsfs}
\usepackage{amsthm}
\DeclareMathOperator{\sech}{sech}
\newtheorem{theorem}{Theorem}
\newtheorem{conjecture}{Conjecture}

\title{Geometry of Massless Scattering in Integrable Superstring}

\author[a]{A. Fontanella}
\author[b]{and A. Torrielli}
\affiliation[a]{Instituto de F\'isica Te\'orica UAM/CSIC \\
C/ Nicol\'as Cabrera, 13–15, C.U. Cantoblanco, \\
E-28049 Madrid, Spain}
\affiliation[b]{Department of Mathematics \\
University of Surrey \\
Guildford, GU2 7XH, UK}
\emailAdd{andrea.fontanella@inv.uam.es}
\emailAdd{a.torrielli@surrey.ac.uk}

\abstract{We consider the action of the $q$-deformed Poincar\'e superalgebra on the massless non-relativistic R-matrix in ordinary (undeformed) integrable $AdS_2 \times S^2 \times T^6$ type IIB superstring theory. 
The boost generator acts non-trivially on the R-matrix, confirming the existence of a non-relativistic rapidity $\gamma$ with respect to which the R-matrix must be of difference form. We conjecture that from a massless AdS/CFT integrable relativistic R-matrix one can obtain the parental massless non-relativistic R-matrix simply by replacing the relativistic rapidity with $\gamma$. We check our conjecture in ordinary (undeformed) $AdS_n \times S^n \times T^{10 - 2n}$, $n = 2, 3$. In the case $n=3$, we check that the matrix part and the dressing factor - up to numerical accuracy for real momenta - obey our prescription. 
In the $n=2$ case, we check the matrix part and propose the non-relativistic dressing factor. 
We then start a programme of classifying R-matrices in terms of connections on fibre bundles. The conditions obtained for the connection are tested on a set of known integrable R-matrices. }


\arxivnumber{1903.10759}

\begin{document}

\maketitle
\flushbottom

\section{Introduction}
\label{sec:level1}

The integrable structures arising in the context of the AdS/CFT correspondence \cite{Beisert:2010jr,Arutyunov:2009ga} provide a very rich testing ground for the interplay between quantum groups and the exact methods of the inverse scattering. This is well demonstrated by superstring theory on the $AdS_3\times S^3\times S^3\times S^1$ and the $AdS_3\times S^3\times T^4$ backgrounds \cite{Babichenko:2009dk,rev3,Borsato:2016hud}. The superconformal algebra underlying the former is the $\mathfrak{D}(2,1;\alpha)\times \mathfrak{D}(2,1;\alpha)$ Lie superalgebra, where $\alpha$ controls the relative radii of the two $S^3$'s; the latter is then obtained by means of an In\"on\"u-Wigner contraction of the former background in the limit $\alpha \to 0$, and displays the superconformal algebra $\mathfrak{psu}(1,1|2)\times\mathfrak{psu}(1,1|2)$. 

The classical integrability of the string sigma-model on these backgrounds was shown in \cite{Babichenko:2009dk,Sundin:2012gc}. The finite-gap equations describing the semi-classical spectrum were given in \cite{OhlssonSax:2011ms}. The excitations are a mixture of massive and massless modes. The massive-massive $S$-matrix  was constructed in \cite{Borsato:2012ud,Borsato:2012ss, Borsato:2013qpa,Borsato:2013hoa} based on a vacuum-preserving algebra composed of a number of centrally-extended  $\mathfrak{psu}(1|1)$ factors. This was succesfully matched against perturbative results \cite{Sundin:2012gc,Rughoonauth:2012qd,Abbott:2012dd,Beccaria:2012kb,
Beccaria:2012pm,Sundin:2013ypa,Bianchi:2013nra, Bianchi:2013nra1, Bianchi:2013nra2}. Scattering involving the {\it massless modes} is a more delicate matter \cite{Sax:2012jv,Lloyd:2013wza}, and it has taken longer to develop a complete world-sheet analysis \cite{BogdanLatest, Borsato:2014hja, Lloyd, Borsato:2015mma,  Abbott:2014rca, MI}. The exact massless scattering theory has been only recently constructed in \cite{Sax:2014mea,Borsato:2016kbm,Borsato:2016xns}. 

At odds with higher dimensional instances of the AdS/CFT correspondence, the dual conformal field theories are harder to grasp. A candidate for the dual to the $AdS_3\times S^3\times T^4$ model was obtained in \cite{Sax:2014mea}, where it was shown to reproduce the predictions of \cite{Sax:2012jv}, although surviving mismatches with perturbation theory \cite{PerLinus} remain unresolved. Comparison between the counting of BPS states emerging from the Bethe ansatz and the number expected from a CFT analysis of the symmetric $T^4$ orbifold point of moduli space $Sym^N (T^4)$ \cite{Eberhardt:2019qcl} was performed in \cite{Baggio:2017kza}. The $AdS_3\times S^3\times S^3\times S^1$ dual CFT has been proposed in \cite{Tong:2014yna}, and the BPS analysis has been revisited in \cite{Baggio:2017kza, Eberhardt:2017fsi}. There, the condition of equal angular momenta on the two $S^3$'s was seen to arise from both the supergravity and the Bethe-equation analysis - see further progress made in \cite{Gaber1,Gaber2,Gaber3,Gaber4,Gaber5,Gaber6,Gaber7}. A host of work on this topic can be found in \cite{Borsato:2015mma,Abbott:2013ixa,Sundin:2013uca,Prin, Prin1, Abbott:2015mla,Abbott:2015mla1,Per,Per1,Per2,Per3,Per4,Per5,Per6,Per7,Per8,Per9,Per10,
Pittelli:2014ria,Regelskis:2015xxa,Hoare:2018jim, Pittelli:2017spf}. In \cite{Bogdan}, the Berenstein-Maldacena-Nastase (BMN) limit of the S-matrix was found to be non-trivial for massless particles and involving purely left-left and right-right scattering. As amply reviewed in \cite{Bogdan}, scattering theory of left-left and right-right moving massless particles in 2D is a non-perturbative phenomenon that Zamolodchikov used in order to describe integrable massless theories at their conformal points. The Thermodynamic Bethe Ansatz (TBA) was derived for such a purely conformal problem and exactly solved to obtain the central charge of the associated CFT to be equal to $6$.

\subsection{The $q$-deformed Poincar\'e superalgebra} 
\label{sec:Theq}

One of the questions concerning the $AdS_3$ superstring massless sector is how much of the relativistic intuition  can be transferred to the superstring setting. When considering the scattering of magnons in the $AdS_5$ setting, \cite{Gomez:2007zr,Charles} reinterpreted the square-root dispersion relation as the Casimir of a $q$-deformed Poincar\'e superalgebra in $1+1$ dimensions. This algebra did not represent the full symmetry of the scattering matrix, nevertheless a boost generator was defined as external to the algebra, and utilised to obtain the known expression for the uniformising rapidity \cite{Beisert:2010jr,Arutyunov:2009ga}. This deformation was a re-casting of the ordinary superstring algebra in a form resembling a trigonometric quantum group. Other ways in which the $q$-Poincar\'e supersymmetries appeared in the $AdS_5$ can be found for example in \cite{Pachol:2015mfa,Kenta, Kenta1, Kenta2}. 

Adopting the same spirit, \cite{Joakim} (and also \cite{Andrea}) demonstrated that a similar $q$-deformed Poincar\'e superalgebra can be written down for $AdS_3$, where it becomes instead a \emph{pseudo-invariance} of the massless $R$-matrix: the algebra excluding the boost is an exact symmetry of the $R$-matrix, while the boost coproduct annihilates it. The $q$-Casimir of this algebra naturally reproduces the massless dispersion relation. This new way of looking at the massless magnons enjoys a series of very natural traits. It allows a compact reformulation of the comultiplication rule, where the coproduct of the momentum determines the one of the other central charges. Interesting connections with phonons and spinons, inspired by \cite{Ballesteros:1999ew}, became quite manifest in this setup. Equally, the boost was employed in \cite{Joakim} to derive a natural uniformising rapidity, capable of reproducing the traditional Zamolodchikov's massless variable in the the relativistic limit. This will be brought to full fruition in this paper and potentially realises the sought-for connection with more traditional relativistic massless scattering. Another related approach was followed in \cite{Riccardo, Riccardo1}.

Following this path, in \cite{Andrea} a new framework was proposed to describe the $AdS_3$ massless sector in terms of a purely geometric framework. The massless $S$-matrix was shown to satisfy a system of differential equations controlled by a flat connection, which was then used to re-write the $S$-matrix  as a path-ordered exponential. Various interpretations were advanced for such an emergent geometry, and the open question of how to connect this to the issue of the dressing phase were left open, although introducing a dressing phase has been understood as a gauge transformation for the connection \cite{Andrea_tesi}.

\subsection{$AdS_2$ and Bethe ansatz}
The $AdS_2 \times S^2 \times T^6$ background \cite{ads2, ads21, ads22, ads23, ads24} is, in this respect, particularly interesting. The holographic dual should either be a superconformal quantum mechanics or a chiral 2D CFT \cite{dual,dual1, dual2, dual3, dual4, gen, gen1, gen2, gen3, gen4, gen5, gen6, gen7, gen8, gen9, gen10, gen11, gen12, gen13, gen14, gen15}. The string sigma model
is formulated on a $\frac{PSU(1,1|2)}{ SO(1,1) \times SO(2)}$ supercoset \cite{Metsaev:1998it, Metsaev:1998it1, Metsaev:1998it2,Borsato:2016zcf}, and it has been demonstrated to be classically integrable \cite{Bena:2003wd,Babichenko:2009dk} up to second order in the fermions \cite{Sorokin:2011rr,Cagnazzo:2011at}.

The exact quantum $S$-matrix was conjectured in \cite{Hoare:2014kma} based on the centrally-extended $\mathfrak{psu}(1|1)^2$ residual symmetry of the BMN vacuum \cite{Berenstein:2002jq,amsw, amsw1, amsw2, amsw3, amsw4}, cf. \cite{beis0, beis01, beis02, beis03, beis04, beis05, beis06}. The $S$-matrix of massive excitations satisfies crossing and unitarity, but determining the dressing factor remains an open problem. Perturbation theory gives reasonable agreement \cite{amsw, amsw1, amsw2, amsw3, amsw4}. The massive magnon representations are {\it long}, while the massless ones are {\it short}. The $S$-matrix for massless modes is obtained as a limit of the massive one \cite{BogdanLatest,Zamol2, Zamol21}, where one has to distinguish between right and left movers. The Yangian symmetry of the problem was elucidated in \cite{Hoare:2014kma,Hoare:2014kmaa}. Matching with perturbation theory is much less clear in the massless sector \cite{amsw, amsw1, amsw2, amsw3, amsw4, MI}, and it deserves further investigation. Massless scattering is in fact quite different from the massive one - see {\it e.g.} \cite{Borsato:2016xns}, and it was understood by Zamolodchikov as a way of describing the renormalisation group flow between critical theories \cite{Zamol2, Zamol21}. 

The lack of relativistic invariance one typically experiences in AdS/CFT integrability adds to the complication \cite{BogdanLatest}. If one attempts to take a relativistic (BMN) limit, this turns out to be trivial for massive modes. Crucially, this is not the case in the massless sector: the relativistic limit is non-trivial between right-right and left-left movers \cite{Andrea2}. The Lie-algebra reduces to ${\cal{N}}=1$ supersymmetry \cite{Fendley:1990cy}, but the S-matrix is rather different. The $S$-matrix retains the maximum number of non-zero entries, as in the XYZ/eight-vertex model \cite{Baxter:1972hz, Baxter:1972hz1} and in typical relativistic ${\cal{N}}=1$ theories \cite{Schoutens,MC}. The transfer matrix does not admit a reference state, making it impossible to apply the algebraic Bethe ansatz \cite{Levkovich-Maslyuk:2016kfv} to find the spectrum and test the proposed Bethe equations of \cite{Sorokin:2011rr}. Integrable systems not admitting a reference state are an area of intense current investigation \cite{Nepo, Nepo1, Nepo2, Nepo3, Nepo4, Nepo5, Nepo6}. The approach of \cite{Andrea2} relied on the so-called {\it free-fermion condition} \cite{MC,Ahn:1993qa} combined with the use of {\it inversion relations} \cite{Zamolodchikov:1991vh}. 

This was further checked in \cite{alea}, where the eigenvalues of the transfer matrix were explicitly calculated up to 5 particles, and a conjecture for the complete massless Bethe ansatz was formulated. Following the ideas of Zamolodchikov, the decoupling of right and left modes ought to indicate that we are in fact describing a critical theory \cite{Bogdan}, whose spectrum should be controlled by the Bethe ansatz we have conjectured\footnote{We thank Diego Bombardelli, Bogdan Stefa\'nski and Roberto Tateo for crucial discussions about this point.}.

Furthermore, the free-fermion condition turns out to be also valid for the massive $AdS_2$ scattering \cite{Hoare:2014kma}, thanks to a particular $\mathfrak{u}(1)$ symmetry of the model \cite{amsw, amsw1, amsw2, amsw3, amsw4, Faddeev:1995nf, Faddeev:1995nf1}. Due to a series of remarkable simplifications, and in spite of the complication of the massive S-matrix entries, the procedure was partially extended to the massive case in \cite{alea}, opening the possibility of obtaining manageable expressions which could be then compared with \cite{Sorokin:2011rr}.

\subsection{Plan of this paper}
This paper is organised as follows. In section \ref{sec:Poincare}, we extend the $\mathfrak{su} (1|1)_c$ superalgebra to the $q$-deformed Poincar\'e superalgebra in $d=2$, and we study the action of the boost generator $\mathcal{J}$ on the R-matrix governing the massless non-relativistic scattering in ordinary (undeformed) integrable $AdS_2 \times S^2 \times T^6$ type IIB superstring theory, which leads us to introduce the new variable $\gamma$. In sections \ref{sec:gamma} and \ref{sec:dressing}, we show that $\gamma$ has the meaning of \emph{non-relativistic rapidity}, and we make the following
\begin{conjecture}
In AdS/CFT, every massless non-relativistic R-matrix governing the scattering of right-right (or left-left)  modes is obtained from the massless relativistic one via the substitution
\begin{equation}
\notag
\theta \rightarrow \gamma \ , \qquad\qquad \gamma \equiv \log \tan \frac{p}{4}  \ .
\end{equation} 
\end{conjecture}
In ordinary (undeformed) $AdS_3 \times S^3 \times T^4$, we show that this conjecture holds for both the matrix part of the R-matrix and the dressing factor, with numerical evidence for real momenta. In $AdS_2 \times S^2 \times T^6$, we show that this prescription works for the matrix part of the R-matrix, and we conjecture the non-relativistic dressing factor. In section \ref{sec:universal}, we assume that every R-matrix depending on one real (or complex) parameter $\theta$ can be written as 
\begin{equation}
\notag
R (\theta) =  \mathscr{P} \exp  \bigg(-\int_0^{\theta} d\tau \Gamma_{\theta}(\tau)  \bigg) \mathcal{A}  \ , 
\end{equation}
where ${\cal{A}}=R(0)$ is a $\theta$-independent matrix, and we study which equations physical unitarity, braiding unitarity and the algebra invariance for the R-matrix imply for $(\Gamma , \mathcal{A})$. Our approach is reminiscent of the one developed in \cite{Maillet1,Maillet2,Maillet3, Knizhnik:1984nr}, however it differs from it.

\section{$q$-deformed Poincar\'e for massless $AdS_2 \times S^2 \times T^6$}
\label{sec:Poincare}
The algebra to consider in $AdS_2 \times S^2 \times T^6$ integrable string background is a central extension of the $\mathfrak{psu} (1|1)$ superalgebra\footnote{We focus here only on one single copy of the algebra and R-matrix (see \cite{Hoare:2014kma}), which is sufficient for all our purposes.}, which will be denoted as $\mathfrak{su}(1|1)_c$. The non-trivial graded commutation relations are
\begin{eqnarray}
\{ \mathcal{Q}, \mathcal{Q} \} = \mathcal{P}, \qquad 
\{ \mathcal{S}, \mathcal{S} \} = \mathcal{K}, \qquad
\{ \mathcal{Q}, \mathcal{S} \} = \mathcal{H},
\end{eqnarray}
where $\mathcal{P}, \mathcal{K}$ and $\mathcal{H}$ are central bosonic generators, $\mathcal{Q}$ and $\mathcal{S}$ are fermionic generators. 

We represent the $\mathfrak{su}(1|1)_c$ generators as $2\times 2$ matrices acting on a pair of boson-fermion $( |\phi \rangle , | \psi \rangle )^T$ as
\begin{eqnarray}
\notag
\mathcal{Q} = \begin{pmatrix}
0 & b \\
a & 0 
\end{pmatrix} \ , \qquad
\mathcal{S} = \begin{pmatrix}
0 & d \\
c & 0 
\end{pmatrix} \ , 
\end{eqnarray}
\begin{eqnarray}
\label{Q_S_H_P_K}
\mathcal{H} = H \begin{pmatrix}
1 & 0 \\
0 & 1 
\end{pmatrix} \ , \qquad
\mathcal{P} = P \begin{pmatrix}
1 & 0 \\
0 & 1 
\end{pmatrix} \ , \qquad
\mathcal{K} = K \begin{pmatrix}
1 & 0 \\
0 & 1 
\end{pmatrix} \ 
\end{eqnarray}
where $a, b, c, d, H, P, K \in \mathbb{C}$ are the \emph{representation parameters}\footnote{For generic values of the mass, the independent ones are only $a, b, c, d$.}.
The (non-relativistic) massless representation is given by the following choice of representation parameters
\begin{eqnarray}
\label{par_massless}
&& a =\alpha \sqrt{h \sin (p / 2) },  \qquad\qquad b = \pm \frac{1}{\alpha}  \sqrt{ h \sin ( p / 2)} \ , \nonumber \\
&& c = \pm \alpha \sqrt{ h \sin ( p / 2) }, \qquad\ \  d = \frac{1}{\alpha} \sqrt{ h \sin ( p / 2)} \ , 
\end{eqnarray}
and
\begin{eqnarray}
\label{par2_massless}
H = \mathcal{E} \ , \qquad P = K = \pm 2 h \sin \frac{p}{2} \ ,  
\end{eqnarray}
where $h$ is the coupling constant, while $\mathcal{E}$ and $p$ stand for the energy and the momentum of the particle. They are related via the dispersion relation  
\begin{eqnarray}
\label{disp_massless_AdS2}
\mathcal{E} = 2  h \, \Big\lvert\sin \frac{ p}{2}\Big\rvert ,
\end{eqnarray} 
which is a shortening condition.
In the formulas above, the upper (lower) sign is associated with right (left) movers\footnote{For the left movers case, one also needs to account for a global factor of $\sqrt{-1} = i$ according to our choice of branch, which matters if one considers the mixed right-left and left-right coproducts.}, and $\mathfrak{Re} p \in [0,\pi]$ for right movers, or $\mathfrak{Re}  p \in [-\pi,0]$ for left movers. 

The non-relativistic massless R-matrix, invariant under the representation (\ref{Q_S_H_P_K}), has been found in \cite{Hoare:2014kma}. 
For the \emph{right-right} scattering, the R-matrix is
\begin{eqnarray}
\label{Rnonrel1}
R^{RR} = \begin{pmatrix}
1 & 0 & 0 &  \pm \frac{1}{\alpha^2} \Big[\frac{\tan \frac{p_1}{4}}{\tan \frac{p_2}{4}}\Big]^{\pm \frac{1}{2}}\\
0 & \pm 1  & \Big[\frac{\tan \frac{p_1}{4}}{\tan \frac{p_2}{4}}\Big]^{\pm \frac{1}{2}} & 0 \\
0 & \Big[\frac{\tan \frac{p_1}{4}}{\tan \frac{p_2}{4}}\Big]^{\pm \frac{1}{2}} & \mp 1 & 0 \\
\pm \alpha^2\Big[\frac{\tan \frac{p_1}{4}}{\tan \frac{p_2}{4}}\Big]^{\pm \frac{1}{2}} & 0 & 0 & -1
\end{pmatrix} \ , 
\end{eqnarray}
and for the \emph{left-left} is
\begin{eqnarray}
\label{Rnonrel2}
R^{LL} = \begin{pmatrix}
1 & 0 & 0 &  \pm \frac{1}{\alpha^2} \Big[\frac{\tan \frac{p_1}{4}}{\tan \frac{p_2}{4}}\Big]^{\mp \frac{1}{2}}\\
0 & \mp 1  & \Big[\frac{\tan \frac{p_1}{4}}{\tan \frac{p_2}{4}}\Big]^{\mp \frac{1}{2}} & 0 \\
0 & \Big[\frac{\tan \frac{p_1}{4}}{\tan \frac{p_2}{4}}\Big]^{\mp \frac{1}{2}} & \pm 1 & 0 \\
\pm \alpha^2\Big[\frac{\tan \frac{p_1}{4}}{\tan \frac{p_2}{4}}\Big]^{\mp \frac{1}{2}} & 0 & 0 & -1
\end{pmatrix}.
\end{eqnarray}
The upper (lower) signs in (\ref{Rnonrel1}) and (\ref{Rnonrel2}) correspond to $f \rightarrow +1$ ( $f \rightarrow -1$) when considering the limit from the massive R-matrix, as explained in \cite{Andrea2}.
In order to understand the dispersion relation (\ref{disp_massless_AdS2}) as the vanishing of a quadratic Casimir, we extend the superalgebra $\mathfrak{su}(1|1)_c$ to the $q$-deformed Poincar\'e superalgebra in $d=2$, $\mathfrak{E}_q (1|1)$. This requires one to introduce the \emph{boost} generator $\mathfrak{J}$, with non trivial commutation relations:
\begin{eqnarray}
\label{qpoincare_algebra}
\notag
&&\{ \mathcal{Q}, \mathcal{Q} \} = \mathcal{P} \ , \qquad
\{ \mathcal{S}, \mathcal{S} \} = \mathcal{K} \ , \qquad
\{ \mathcal{Q}, \mathcal{S} \} = \mathcal{H} \ , \\
\notag
&& [ \mathcal{J} , \mathcal{Q} ] = \frac{i}{2 \sqrt{\mu}} \frac{e^{i \frac{p}{2}} + e^{- i \frac{p}{2}}}{2} \mathcal{Q} \ , \qquad\quad\ \,
[\mathcal{J} , \mathcal{S} ] = \frac{i}{2 \sqrt{\mu}} \frac{e^{i\frac{p}{2}} + e^{- i \frac{p}{2}}}{2} \mathcal{S}  \ , \\
\notag
&& [ \mathcal{J}, \mathcal{P}]  = \frac{i}{\sqrt{\mu}} \frac{e^{i\frac{p}{2}} + e^{- i\frac{p}{2}}}{2} \mathcal{P} \ , \qquad\qquad
 [ \mathcal{J}, \mathcal{K}]  = \frac{i}{\sqrt{\mu}} \frac{e^{i\frac{p}{2}} + e^{- i\frac{p}{2}}}{2} \mathcal{K} \ , \\
&& [ \mathcal{J}, \mathcal{H} ] =  \frac{e^{ip} - e^{-i p}}{2 \mu} \mathds{1} \ , \qquad\qquad\qquad
[\mathcal{J} , p] = i \mathcal{H} \ , 
\end{eqnarray}
where $\mu \equiv h^{-2}$ and the deformation parameter $q$ is related to the coupling constant $h$ via
\begin{equation}
\log q = \frac{i}{h^2} \ . 
\end{equation}
The representation of the boost on a single particle state is $\mathcal{J} = i \mathcal{H} \partial_p$. 
The coproducts for the generators of $\mathfrak{E}_q (1|1)$ are
\begin{eqnarray}
\notag
&&\Delta (\mathcal{Q}) = \mathcal{Q} \otimes e^{ i \frac{p}{4}} + e^{- i \frac{p}{4}} \otimes Q \ , \qquad\qquad
\Delta (\mathcal{S}) = \mathcal{S} \otimes e^{ i \frac{p}{4}} + e^{- i \frac{p}{4}} \otimes S \ , \\ 
\notag
&& \Delta (\mathcal{P}) = \mathcal{P} \otimes {e^{i \frac{p}{2}}} + {e^{-i \frac{p}{2}}} \otimes \mathcal{P} \ , \qquad\qquad
\Delta (\mathcal{K}) = \mathcal{K} \otimes {e^{i \frac{p}{2}}} + {e^{-i \frac{p}{2}}} \otimes \mathcal{K} \ , \\
&&\Delta(\mathcal{H})=  \mathcal{H} \otimes {e^{i \frac{p}{2}}} + {e^{-i \frac{p}{2}}} \otimes \mathcal{H} \ , \qquad\quad\ \ 
\Delta (\mathcal{J}) = \mathcal{J} \otimes e^{- i \frac{p}{2}} +  e^{i \frac{p}{2}} \otimes \mathcal{J}   \ .
\end{eqnarray}
It is very interesting that the commutation relations (\ref{qpoincare_algebra}) indicate how the boost operator has a similar action to the one of the outer automorphism $\cal{D}$ of the centrally-extended $\mathfrak{psu}(1|1)$ superalgebra\footnote{Where the odd elements $\mathcal{Q}, \mathcal{S}$ have weight $1/2$ (e.g.  $[\mathcal{D}, \mathcal{Q}] = \frac{1}{2} \mathcal{Q}$) and even elements $\mathcal{P}, \mathcal{K}, \mathcal{H}$ have weight $1$ (e.g.  [$\cal{D}, \mathcal{P}] = \mathcal{P}$).}, although they are not the same. A first difference partly resides in the momentum-dependent proportionality factor, which effectively deforms the right hand side of the commutation relations.  This is particularly clear when $p$ is promoted to a \emph{generator} in a universal ({\it i.e.} representation-independent) reformulation of the $\mathfrak{E}_q (1|1)$ superalgebra. A second main difference is the non-standard coproduct for the boost operator, which signals a non-locality in its two-particle action.

The R-matrices $R^{RR}$ and $R^{LL}$ (\ref{Rnonrel1}) and (\ref{Rnonrel2}) are invariant under the action of $\mathcal{Q}$, $\mathcal{S}$ and the central bosonic generators, i.e.
\begin{equation}
\Delta^{op} (\mathfrak{a}) R = R \Delta (\mathfrak{a}) \ , \qquad\qquad
\mathfrak{a} = \mathcal{Q}, \mathcal{S}, \mathcal{P}, \mathcal{K}, \mathcal{H}\ , 
\end{equation}
however, they are \emph{not} invariant under the boost action $\Delta(J)$. Moreover, they are neither annihilated by $\Delta(J)$ nor by $\Delta^{op}(J)$, in contrast to the boost action on the massless R-matrix in $AdS_3 \times S^3 \times T^4$ discussed in \cite{Andrea}.  
Nevertheless, the R-matrix $R^{RR}$ satisfies for instance the following condition  
\begin{equation}
\label{JJR}
\bigg( w \Delta(\mathcal{J}) + w^{op} \Delta^{op} (J) \bigg) R = 0 \ , 
\end{equation}
where
\begin{equation}
w = e^{-\frac{i}{4} ( p \otimes \mathds{1} - \mathds{1} \otimes p)} \ , \qquad
w^{op} =  e^{-\frac{i}{4} (\mathds{1} \otimes p -  p \otimes \mathds{1})}\ . 
\end{equation}
If we introduce the new variable $\gamma$ - cf. \cite{Joakim} - as
\begin{equation}
\label{gamma}
\gamma \equiv \log \tan \frac{p}{4} \ , 
\end{equation}
then equation (\ref{JJR}) becomes
\begin{equation}
\label{difference_gamma}
\bigg( \frac{\partial}{\partial \gamma_1} + \frac{\partial}{\partial \gamma_2} \bigg) R = 0 \ , 
\end{equation}
which implies that $R$ depends only on the difference $\gamma_1 - \gamma_2$. In the relativistic limit
\begin{equation}
\label{rel_limit}
p \rightarrow \epsilon q\,  \equiv \epsilon e^{\theta}\ , \qquad\qquad
h \rightarrow \frac{c}{\epsilon} \ , 
\end{equation}
where $\epsilon \rightarrow 0$, and $\theta$ is the rapidity of the particle, we have that equation (\ref{difference_gamma}) becomes
\begin{equation}
\label{JJR_relativ}
\bigg( \frac{\partial}{\partial \theta_1} + \frac{\partial}{\partial \theta_2} \bigg) R^{rel} = 0 \ , 
\end{equation}
which states that the relativistic R-matrix $R^{rel}$ only depends on $\theta \equiv \theta_1 - \theta_2$. 

We can provide a second possible choice of coproduct, which is a homomorphism for the Borel-type subalgebra of the $\mathfrak{E}_q (1|1)$ superalgebra formed by the generators $\mathcal{Q}$, $\mathcal{P}$ and $\mathcal{J}$ where $\mathcal{Q}$ is a single real supercharge\footnote{For our specific choice of representation the generators $\mathcal{Q}$ and $\mathcal{S}$ coincide in the boson-fermion representation, so they can be used to think of a very small algebra controlling the scattering problem where they appear as a unique generator. The same applies to $\mathcal{K}, \mathcal{P}$ and $\mathcal{H}$.} This coproduct satisfies the coassociativity property, and is given by
\begin{eqnarray}
\label{coproduct_hat}
\notag
&&\hat{\Delta} (\mathcal{Q}) = \mathcal{Q} \otimes e^{ i \frac{p}{4}} + e^{- i \frac{p}{4}} \otimes Q \ , \qquad\qquad \hat{\Delta}(\mathcal{P})=  \mathcal{P} \otimes {e^{i \frac{p}{2}}} + {e^{-i \frac{p}{2}}} \otimes \mathcal{P} \ , \\
&&\hat{\Delta} (\mathcal{J}) = \mathcal{J} \otimes e^{i \frac{p}{2}} +  e^{-i \frac{p}{2}} \otimes \mathcal{J}  + \frac{1}{2} e^{- i \frac{p}{4}} \mathcal{Q} \otimes e^{i \frac{p}{4}} \mathcal{Q}  \ .
\end{eqnarray}
The coproduct $\hat{\Delta}$ differs from $\Delta$ only for the boost generator $\mathcal{J}$, otherwise they are the same. We find again that the R-matrices (\ref{Rnonrel1}) and (\ref{Rnonrel2}) are not invariant under $\hat{\Delta} (\mathcal{J})$, and neither are they annihilated by $\hat{\Delta} (\mathcal{J})$ and $\hat{\Delta}^{op} (\mathcal{J})$. 
However the following combination annihilates the R-matrix 
\begin{equation}
\label{JJR_hat}
\bigg( z \hat{\Delta}(\mathcal{J}) + z^{op} \hat{\Delta}^{op} (\mathcal{J}) \bigg) R = 0 \ , 
\end{equation}
where $z = w^{-1}$, i.e.
\begin{equation}
z = e^{\frac{i}{4} ( p \otimes \mathds{1} - \mathds{1} \otimes p)} \ , \qquad\qquad
z^{op} = e^{\frac{i}{4} ( \mathds{1} \otimes p - p \otimes \mathds{1})} \ . 
\end{equation}
One can introduce the variable $\gamma$ defined as in (\ref{gamma}) and equation (\ref{JJR_hat}) becomes (\ref{difference_gamma}), which
in the relativistic limit reproduces (\ref{JJR_relativ}). 

We shall comment how the introduction of the dressing factor $\Phi$ affects equation (\ref{difference_gamma}). In a similar spirit to adding a $\mathfrak{u}(1)$ part to a would-be connection, as it was shown for the $AdS_3$ case in \cite{Andrea}, we have that the analogous of (\ref{difference_gamma}) for the dressed R-matrix, \emph{i.e.}
\begin{equation}
\bigg( \frac{\partial}{\partial \gamma_1} + \frac{\partial}{\partial \gamma_2} \bigg) \tilde{R} = 0 \ , \qquad\qquad
\tilde{R} \equiv \Phi R \ . 
\end{equation}
implies that 
\begin{equation}
R \bigg( \frac{\partial}{\partial \gamma_1} + \frac{\partial}{\partial \gamma_2} \bigg) \Phi + \Phi \bigg( \frac{\partial}{\partial \gamma_1} + \frac{\partial}{\partial \gamma_2} \bigg) R = 0
\end{equation}
and by using (\ref{difference_gamma}) for the undressed R-matrix, this in turns implies that 
\begin{equation}
\bigg( \frac{\partial}{\partial \gamma_1} + \frac{\partial}{\partial \gamma_2} \bigg) \Phi = 0 \ , 
\end{equation}
\emph{i.e.} the dressing factor must depend only on the difference $\gamma_1 - \gamma_2$. We shall see in section \ref{sec:dressing} that this condition is indeed satisfied by the dressing factor for the $AdS_3$ case. This means that also the dressed R-matrix must only depend on $\gamma_1 - \gamma_2$.

\section{$\gamma$ as a non-relativistic rapidity}
\label{sec:gamma}
The variable $\gamma$ defined in (\ref{gamma}) emerges from the boost action on the R-matrix via equations (\ref{JJR}) and (\ref{JJR_hat}). The domain of $\gamma$ is $(- \infty , 0]$, which is a consequence of the fact that the domain of $p$ for a right mover particle is $[0, \pi]$. 
The energy (\ref{disp_massless_AdS2}) in terms of the variable $\gamma$ is
\begin{equation}
\mathcal{E} =  \frac{2 h}{\cosh  \gamma } \ ,  
\end{equation}
and the group velocity $v_g$ 
\begin{equation}
v_g = \frac{d \mathcal{E}}{d p} = - \frac{h}{\tanh  \gamma} \ . 
\end{equation}
In the relativistic limit (\ref{rel_limit}), the variable $\gamma$ tends to 
\begin{equation}
\gamma \rightarrow \theta +  \log \frac{\epsilon}{4} \ , 
\end{equation}
which diverges logarithmically. However in the context of R-matrices, equation  (\ref{JJR_relativ}) tells us that the R-matrix depends only on the difference $\gamma_1 - \gamma_2$, which is a well defined variable in the relativistic limit. 
From now on, we shall denote $\gamma \equiv \gamma_1 - \gamma_2$. 

In terms of the variable $\gamma$, the R-matrices (\ref{Rnonrel1}) and (\ref{Rnonrel2}) becomes 
\begin{eqnarray}
\label{Rnonrel1_gamma}
R^{RR} = \begin{pmatrix}
1 & 0 & 0 &  \pm \frac{1}{\alpha^2} e^{\pm \frac{\gamma}{2}}\\
0 & \pm 1  & e^{\pm \frac{\gamma}{2}} & 0 \\
0 & e^{\pm \frac{\gamma}{2}} & \mp 1 & 0 \\
\pm \alpha^2 e^{\pm \frac{\gamma}{2}}  & 0 & 0 & -1
\end{pmatrix} \ , 
\end{eqnarray}
and 
\begin{eqnarray}
\label{Rnonrel2_gamma}
R^{LL} = \begin{pmatrix}
1 & 0 & 0 &  \pm \frac{1}{\alpha^2} e^{\mp \frac{\gamma}{2}} \\
0 & \mp 1  & e^{\mp \frac{\gamma}{2}} & 0 \\
0 & e^{\mp \frac{\gamma}{2}} & \pm 1 & 0 \\
\pm \alpha^2 e^{\mp \frac{\gamma}{2}} & 0 & 0 & -1
\end{pmatrix} \ .
\end{eqnarray}
The relativistic massless R-matrices to which (\ref{Rnonrel1_gamma}) and (\ref{Rnonrel2_gamma}) tend in the relativistic limit are encoded in \emph{Solution 3} and \emph{Solution 5} studied in \cite{Andrea2}, and they are respectively\footnote{We observe that \emph{Solution 3} can be rewritten in terms of three $2\times 2$ matrices which satisfy the quaternion algebra, as explained in Appendix \ref{appx:quaternionic}.}
\begin{eqnarray}
R^{RR}_{rel} = \begin{pmatrix}
1 & 0 & 0 &  \pm \frac{1}{\alpha^2} e^{\pm \frac{\theta}{2}}\\
0 & \pm 1  & e^{\pm \frac{\theta}{2}} & 0 \\
0 & e^{\pm \frac{\theta}{2}} & \mp 1 & 0 \\
\pm \alpha^2 e^{\pm \frac{\theta}{2}}  & 0 & 0 & -1
\end{pmatrix} \ ,  
\end{eqnarray}
and 
\begin{eqnarray}
R^{LL}_{rel} = \begin{pmatrix}
1 & 0 & 0 &  \pm \frac{1}{\alpha^2} e^{\mp \frac{\theta}{2}} \\
0 & \mp 1  & e^{\mp \frac{\theta}{2}} & 0 \\
0 & e^{\mp \frac{\theta}{2}} & \pm 1 & 0 \\
\pm \alpha^2 e^{\mp \frac{\theta}{2}} & 0 & 0 & -1
\end{pmatrix} \ . 
\end{eqnarray}
We notice that, if we knew only the \emph{relativistic} massless R-matrices, we could construct the parental \emph{non-relativistic} massless R-matrices simply by replacing 
\begin{equation}
\label{substituion}
\theta \equiv \theta_1 - \theta_2 \  \longrightarrow \  \gamma \equiv  \gamma_1 - \gamma_2 \ . 
\end{equation}
Interestingly, this prescription also works for the R-matrix in ordinary (undeformed) $AdS_3 \times S^3 \times T^4$. 
The relativistic massless R-matrix in this background is \cite{Bogdan}
\begin{eqnarray}
R_{AdS_3}^{rel} = 
\begin{pmatrix}
1 & 0 & 0 & 0 \\
0 &  - \tanh \frac{\theta}{2} & \sech \frac{\theta}{2} & 0 \\
0 & \sech \frac{\theta}{2} & \tanh \frac{\theta}{2} & 0 \\
0 & 0 & 0 & -1 
\end{pmatrix} \ , 
\end{eqnarray}
and the non-relativistic massless R-matrix in \cite{BogdanLatest}, expressed in terms of the $\gamma$ variable, is 
\begin{eqnarray}
\label{AdS3_non_rel}
R_{AdS_3}^{non-rel} = 
\begin{pmatrix}
1 & 0 & 0 & 0 \\
0 &  - \tanh \frac{\gamma}{2} & \sech \frac{\gamma}{2} & 0 \\
0 & \sech \frac{\gamma}{2} & \tanh \frac{\gamma}{2} & 0 \\
0 & 0 & 0 & -1 
\end{pmatrix} \ . 
\end{eqnarray}
This suggests once again that the non-relativistic R-matrix can be obtained from the relativistic one simply by the substitution (\ref{substituion}). 

The variable $\gamma$ can be thought of as a \emph{non-relativistic rapidity}. We are naturally brought to conjecture that this simple relationship between non-relativistic and relativistic massless AdS/CFT R-matrices, where one simply replaces $\theta = \theta_1 - \theta_2$ with 
\begin{equation}
\gamma = \gamma_1 - \gamma_2 = \log \tan \frac{p_1}{4} - \log \tan \frac{p_2}{4},
\end{equation} 
will extend in more generality, as we shall now further confirm.

\section{The dressing factor}
\label{sec:dressing}

\subsection{$AdS_3 \times S^3 \times T^4$}

Our intuition is reinforced by considerations on the dressing factor, which the R-matrix needs to be equipped with. In \cite{Borsato:2016xns} this dressing factor was derived by solving the massless non-relativistic crossing equation, while in \cite{Bogdan} the relativistic limit was obtained and shown to coincide with Zamolodchikov's dressing factor for Sine-Gordon theory at a special value of the coupling. According to the expectations developed in the previous section, the same Zamolodchikov's analytic form should now hold and reproduce the  expression derived in \cite{Borsato:2016xns}, simply by replacing 
\begin{equation}
\theta = \theta_1 - \theta_2 \qquad \mbox{by} \qquad \gamma = \gamma_1 - \gamma_2. 
\end{equation}
This is in part because the matrix part of the R-matrix determines the r.h.s. of the crossing equation, which shall therefore be identical to the relativistic one with $\gamma$ replacing $\theta$. In particular, this should reveal how the dressing factor of \cite{Borsato:2016xns} is in fact of difference form in the variable $\gamma$.

We have considered the Hernandez-Lopez (HL) term and we have numerically verified that this is indeed the case for real momenta, by using the dilogarithm expression of the HL contribution to the dressing phase provided in \cite{Beisert:2006ib} evaluated for massless variables $x_i^\pm = e^{\pm i \frac{p_i}{2}}$, $i=1,2$:
\begin{eqnarray}
&&\chi(x,y) \equiv \frac{1}{2\pi} \bigg[-\mbox{Li}_2 \frac{\sqrt{x} - \frac{1}{\sqrt{y}}}{\sqrt{x} - \sqrt{y}}
-\mbox{Li}_2 \frac{\sqrt{x} + \frac{1}{\sqrt{y}}}{\sqrt{x} + \sqrt{y}}+\mbox{Li}_2 \frac{\sqrt{x} - \frac{1}{\sqrt{y}}}{\sqrt{x} + \sqrt{y}}+\mbox{Li}_2 \frac{\sqrt{x} + \frac{1}{\sqrt{y}}}{\sqrt{x} - \sqrt{y}}\bigg]\,,\nonumber\\\nonumber\\
&& \Theta = \chi(x_1^+,x_2^+) + \chi(x_1^-,x_2^-)-\chi(x_1^-,x_2^+)-\chi(x_1^+,x_2^-) - (1 \leftrightarrow 2),\nonumber\\\nonumber\\
&&x_i^\pm = e^{\pm i \frac{p_i}{2}}, \qquad i=1,2, \qquad \gamma = \gamma_1 - \gamma_2, \qquad \gamma_i = \log \tan \frac{p_1}{4},
\end{eqnarray}
and (numerically\footnote{The numerics is done primarily for real momenta - although we have obtained numerical evidence that there exists a double-sided interval of certain regions of the real line where our statements do extend to the complex plane - and the functions are highly oscillatory. It would therefore be highly desirable to have an analytic proof in order to reach a conclusive statement.} up to $\sim 10^{-15}$)
\begin{eqnarray} 
&&\Big(\sin \frac{p_1}{2} \frac{\partial}{\partial p_1} + \sin \frac{p_2}{2} \frac{\partial}{\partial p_2}\Big) \Theta = \frac{\partial}{\partial (\gamma_1 + \gamma_2)} \Theta=0, \\ 
&&\qquad \qquad \qquad \Big(\sin \frac{p_1}{2} \frac{\partial}{\partial p_1} - \sin \frac{p_2}{2} \frac{\partial}{\partial p_2}\Big) \Theta = \frac{d}{d\gamma} \Theta = \frac{4 \gamma}{\sinh \gamma}.
\end{eqnarray}
Not only the HL term of the dressing factor only depends in massless kinematics - with the above numerical {\it proviso} - on the difference $\gamma_1 - \gamma_2$, but it satisfies - again to the numerical accuracy - the relation (4.14) in \cite{Bogdan} - hallmark of the fact
that the factor is provided in fact by expression (5.36) in \cite{Bogdan}. 

It is amusing to notice that a natural consequence of this observation, if corroborated by analytic evidence and extended to complex values of momenta, would be that the massless non-relativistic dressing factor is meromorphic in the complex plane $\gamma$, with only poles on the imaginary axis and no pole in the region $\mathfrak{Im} \gamma \in [0,\pi]$, and would clearly possess all the attributes of a relativistic phase when considered in the new variable. 

\smallskip

We remark that it would have been rather difficult to spot all these remarkable facts, had it not been for the realisation that the boost operator in the new form provided in this paper might act on the dressing factor as well and in fact annihilates it. 

\subsection{$AdS_2 \times S^2 \times T^6$}
In \cite{Andrea2}, the dressing factor for the relativistic massless R-matrix, dubbed \emph{Solution 3} in that paper, has been found, and an integral representation is given in Appendix \ref{appx:integral}. Supported by the check that our conjecture works also for the dressing factor in the $AdS_3$ case, we infer that the dressing factor for the massless non-relativistic $AdS_2$ case is 
\begin{equation}
\Omega (\gamma ) = B \exp A (\gamma) \ , 
\end{equation}
where
\begin{equation}
B = \frac{e^{- i \frac{\pi}{8}}}{\sqrt{2}} \ , \qquad\qquad
A(\gamma) = \frac{\gamma}{4} + \frac{1}{2} \int_0^\infty \frac{dx}{x} \frac{\cosh [x (1 - \frac{2 \gamma}{i \pi})] - \cosh x}{\cosh x \, \sinh 2 x} \ .
\end{equation}
We checked that this conjectured solution for the dressing phase satisfies the non-relativistic crossing equation derived in \cite{alea}, which reads in terms of the momenta $p_1, p_2$ as follows
\begin{equation}
\Omega(p_1, p_2) \Omega(\bar{p}_1, p_2) = \frac{\sin\frac{p_1}{4}\cos\frac{p_2}{4}}{\sin\frac{p_1+p_2}{4}} \ . 
\end{equation}

\section{Universal R-matrix and Connections}
\label{sec:universal}
In \cite{Andrea}, a geometric interpretation has been found for the R-matrix in $AdS_3 \times S^3 \times T^4$, see also \cite{Andrea_tesi} for a review. 
As a consequence of the pseudo-invariance under the boost generators of $\mathfrak{E}_q (1|1)_L \oplus \mathfrak{E}_q (1|1)_R$, the non-relativistic massless R-matrix (\ref{AdS3_non_rel}) must satisfy the following parallel condition 
\begin{equation}
\label{DR=0}
D_M R \equiv \bigg[\frac{\partial}{\partial p_M} + \Gamma^{AdS_3}_M\bigg]R = 0, \qquad\qquad (M =1,2) \ ,
\end{equation}
where
\begin{eqnarray}
\label{connection}
&& \Gamma^{AdS_3}_1 \equiv -\frac{1}{4} \sqrt{\frac{\sin \frac{p_2}{2}}{\sin \frac{p_1}{2}}} \, \frac{Y}{\sin \frac{p_1 + p_2}{4}} \ ,\nonumber \\
&& \Gamma^{AdS_3}_2 \equiv \frac{1}{4} \sqrt{\frac{\sin \frac{p_1}{2}}{\sin \frac{p_2}{2}}} \, \frac{Y}{\sin \frac{p_1 + p_2}{4}} \ , 
\end{eqnarray} 
and
\begin{equation} 
Y \equiv \big[ \mathsf{E}_{12} \otimes \mathsf{E}_{21} + \mathsf{E}_{21} \otimes \mathsf{E}_{12}\big]\ , \qquad
\mathsf{E}_{12} \equiv \begin{pmatrix}0&1\\0&0\end{pmatrix}, \qquad \mathsf{E}_{21} \equiv \begin{pmatrix}0&0\\1&0\end{pmatrix} \ . 
\end{equation} 
This is turns implies that the R-matrix (\ref{AdS3_non_rel}) can be written as 
\begin{equation}
\label{int_R}
R \big[\gamma (\lambda)\big] = \Pi \circ \mathscr{P} \exp  \bigg(\int_{\gamma(0)}^{\gamma(\lambda)} dp^M \Gamma^{AdS_3}_M \bigg) \  , 
\end{equation}
where $\mathscr{P} \exp$ is the path-ordered exponential, and $\gamma(0) = (p, p)$ with $p$ generic, such that for $\lambda = 0$ we obtain $R(p, p) = \Pi$, as can now be easily seen from (\ref{AdS3_non_rel}). 

The relativistic limit (\ref{rel_limit}) of the parallel condition (\ref{DR=0}) has been discussed in \cite{Bogdan}. One obtains that the two equations contained in (\ref{DR=0}) boil down to just one equation:
\begin{equation}
\label{DR=0_rel}
\bigg( \frac{\partial}{\partial \theta} - \frac{Y}{2\cosh \frac{\theta}{2}} \bigg) R(\theta) = 0 \ . 
\end{equation}
This suggests that in the relativistic limit one can make the following replacement of the connection
\begin{eqnarray}
D_M \longrightarrow \frac{D_{\theta}}{\epsilon} \ , \qquad\qquad\qquad
D_{\theta} \equiv \frac{\partial}{\partial \theta} + \Gamma^{AdS_3}_{\theta}\ ,  \qquad \Gamma^{AdS_3}_{\theta} = - \frac{Y}{2\cosh \frac{\theta}{2}} \ , 
\end{eqnarray}
and the base space $T^2$ shrinks to $S^1$ for real momenta - while a suitable complexification of the base space needs to be considered for instance to discuss crossing symmetry. 
Again, by integrating (\ref{DR=0_rel}) between $[0, \theta]$, we obtain
\begin{equation}
R (\theta) = \Pi \circ \mathscr{P} \exp  \bigg(\int_0^{\theta} d\tau \Gamma^{AdS_3}_{\theta}(\tau)  \bigg) \  .
\end{equation}
In the relativistic limit the R-matrix depends only on one parameter $\theta$ instead of two parameters $(p_1, p_2)$. This suggests to first investigate universal properties of the connection when the base space of the fibre bundle has only one (real or complex) dimension.

\subsection{System of equations for $(\Gamma , \mathcal{A})$}
We assume that a generic R-matrix, which depends only on one rapidity-difference variable, can be written as follows
\begin{equation}
\label{R_conj}
R (\theta) =  \mathscr{P} \exp  \bigg(-\int_0^{\theta} d\tau \Gamma_{\theta}(\tau)  \bigg) \mathcal{A}  \ , 
\end{equation}
where
\begin{equation}
\mathcal{A} \equiv R (0) \ .
\end{equation}
We shall \emph{not} assume any property for $\Gamma_{\theta}$ and $\mathcal{A}$, but we shall find a set of constraints by imposing some of the fundamental equations for the R-matrix. 
Note that the parallel condition
\begin{equation}
\label{parallel_generic_gamma}
\bigg( \frac{\partial}{\partial \theta} + \Gamma_{\theta} \bigg) R(\theta) = 0 \ , 
\end{equation}
where $\Gamma_{\theta}$ is now a generic connection, is automatically satisfied by (\ref{R_conj}) with our choice of ordering of the path-ordering exponential.

\subsubsection{Physical unitarity}
The R-matrix must satisfy physical unitarity, which reads
\begin{equation}
\label{phys_unit}
R(\theta) R^{\dagger}(\theta) = \mathds{1} \otimes \mathds{1} \ ,  
\end{equation}
and it is valid for any real value of $\theta$. 
If $R$ is represented in terms of matrices, $^{\dagger}$ is the standard matrix transposition and complex conjugation of its entries. For $\theta = 0$, equation (\ref{phys_unit}) reads
\begin{equation}
\label{phys_unit_A}
\mathcal{A} \mathcal{A}^{\dagger} = \mathds{1} \otimes \mathds{1} \ . 
\end{equation} 
Differentiating equation (\ref{phys_unit}) with respect to $\theta$, we obtain 
\begin{equation}
\Gamma (\theta) = - \Gamma^{\dagger} (\theta) \ , 
\end{equation}
i.e. $\Gamma$ must be anti-hermitian. 

\subsubsection{Braiding unitarity}
Fist we shall derive the parallel equations which $R^{op} (- \theta )$ must satisfy. We recall that $R^{op} \equiv \Pi\,  R\,  \Pi$, where in this picture $\Pi$ acts on states of the representation, and the momenta of the two particles 1 and 2 must be exchanged by hand. 
We first apply the operator $\Pi$ to (\ref{parallel_generic_gamma}) and obtain
\begin{equation}
\bigg( \frac{\partial}{\partial \theta} + \Pi \, \Gamma_{\theta} ( \theta ) \, \Pi \bigg) \Pi \, R (\theta ) \, \Pi =  0 \ . 
\end{equation}
Then we exchange $\theta \rightarrow \tilde{\theta} \equiv - \theta$ and obtain 
\begin{equation}
\bigg( \frac{\partial}{\partial \tilde{\theta}} + \Pi \, \Gamma_{\theta} ( \tilde{\theta} ) \, \Pi \bigg) R^{op} (\tilde{\theta} ) =  0 \ , 
\end{equation}
which can be integrated between $0$ and $\tilde{\theta}$ to obtain
\begin{equation}
R^{op} (\tilde{\theta} ) =  \mathscr{P} \exp \bigg(- \int_0^{\tilde{\theta}} d \tau \Pi \, \Gamma_{\theta} ( \tau ) \, \Pi \bigg) \Pi \, \mathcal{A}	 \, \Pi   \ ,
\end{equation}
where we imposed the initial condition
\begin{equation}
R^{op} (0) = \Pi \, R (0) \, \Pi = \Pi \, \mathcal{A} \, \Pi \ . 
\end{equation}
Finally, we exchange $\tau \rightarrow - \tau$ and obtain
\begin{equation}
R^{op} (-\theta) =   \mathscr{P} \exp \bigg( \int_0^{\theta} d \tau \Pi \, \Gamma_{\theta} ( - \tau ) \, \Pi \bigg) \Pi \, \mathcal{A} \, \Pi  \ . 
\end{equation}
The braiding unitarity equation 
\begin{equation}
\label{braiding_unit}
R (\theta) R^{op} (- \theta ) = \mathds{1} \otimes \mathds{1} \ , 
\end{equation}
imposes a condition on $\mathcal{A}$ and $\Gamma_{\theta}$, which reads
\begin{equation}
\label{braidingU_gamma}
 \mathscr{P} \exp \bigg(  - \int_0^{\theta} d \tau  \Gamma_{\theta} (  \tau )  \bigg) \mathcal{A} \,
  \mathscr{P} \exp \bigg(  \int_0^{\theta} d \tau  \Gamma^{op}_{\theta} ( - \tau )  \bigg)  \mathcal{A}^{op} = \mathds{1} \otimes \mathds{1} \ , 
\end{equation}
where we have defined
\begin{equation}
\Gamma^{op}_{\theta} (\theta) \equiv \Pi \,\Gamma_{\theta} (\theta)\, \Pi \ , \qquad\qquad
\mathcal{A}^{op} \equiv \Pi \, \mathcal{A} \, \Pi \ . 
\end{equation}
For $\theta = 0$, equation (\ref{braidingU_gamma}) reduces to 
\begin{equation}
\label{braid_unit_A}
\mathcal{A} \,  \mathcal{A}^{op}  = \mathds{1} \otimes \mathds{1} \ . 
\end{equation}
Two obvious solutions to equation (\ref{braid_unit_A}) are\footnote{All the R-matrices we are aware of do follow $\mathcal{A} = \Pi$, cf. condition (1.15) in \cite{Tarasov:1983cj}.} 
\begin{equation}
\mathcal{A} = \Pi \ , \qquad\qquad
\mathcal{A} = \mathds{1} \ .  
\end{equation}
Comparing physical unitarity (\ref{phys_unit}) with braiding unitarity (\ref{braiding_unit}), we also have that 
\begin{equation}
R^{op}(- \theta ) = R^{\dagger}(\theta) \ , \qquad\qquad
\mathcal{A}^{op} = \mathcal{A}^{\dagger} \ . 
\end{equation}

\noindent $\bullet$ \textbf{Case $\mathcal{A} = \Pi$}

\noindent Equation (\ref{braidingU_gamma}) becomes
\begin{equation}
\label{braiding_u_A=P}
\mathscr{P} \exp \bigg( - \int_0^{\theta} d \tau \Gamma_{\theta} (\tau) \bigg) \mathscr{P} \exp \bigg(\int_0^{\theta} d \tau \Gamma_{\theta} (-\tau) \bigg) = \mathds{1} \otimes \mathds{1} \ , 
\end{equation}
The theorem in Appendix \ref{useful_th}, with $A(\tau) = i  \Gamma_{\theta} (\tau)$ and $B(\tau) = i \Gamma_{\theta} (- \tau)$, implies that
\begin{equation}
\Gamma_{\theta} (\theta) = \Gamma_{\theta} (- \theta) \ , 
\end{equation}
i.e. $\Gamma_{\theta} (\theta)$ must be an \emph{even} function. 
We also have the following:

\begin{theorem}
Continuous deformations of the solution $\mathcal{A} = \Pi$ are not solutions.
\end{theorem}
\begin{proof}
Suppose that continuous deformations of the solution $\mathcal{A} = \Pi$ are still solutions to (\ref{braid_unit_A}). We write the generic deformed solution as 
\begin{equation}
\label{deformed_A}
\mathcal{A} = \Pi + \epsilon \, \delta \mathcal{A} + \mathcal{O}(\epsilon^2)\ , 
\end{equation}
where $\epsilon \in \mathbb{R}$ is an arbitrary small real parameter, and $\delta \mathcal{A} \in \mathcal{U}[\mathfrak{g}] \otimes \mathcal{U}[\mathfrak{g}]$ is the deformation. We shall neglect terms which are higher order powers in $\epsilon$. 
If we impose (\ref{deformed_A}) to be a solution to (\ref{braid_unit_A}), we obtain 
\begin{equation}
\big[\Pi + \epsilon \, \delta \mathcal{A} + \mathcal{O}(\epsilon^2)\big] \Pi \big[\Pi + \epsilon \, \delta \mathcal{A} + \mathcal{O}(\epsilon^2)\big] \Pi = \mathds{1} \otimes \mathds{1} \ , 
\end{equation}
which at zeroth order in $\epsilon$ gives us
\begin{equation}
\Pi^4= \mathds{1} \otimes \mathds{1} \ , 
\end{equation}
and at first order in $\epsilon$
\begin{equation}
2 \delta\mathcal{A} \, \Pi = 0 \ , 
\end{equation}
After multiplying the equation above on the right by $\Pi$, we obtain 
\begin{equation}
\delta \mathcal{A} = 0 \ , 
\end{equation}
i.e. a continuous deformation of the solution $\mathcal{A} = \Pi$ is not a solution to (\ref{braid_unit_A}). 

\end{proof}

\noindent $\bullet$ \textbf{Case $\mathcal{A} = \mathds{1}$}

\noindent Equation (\ref{braidingU_gamma}) becomes
\begin{equation}
\label{braiding_u_A=Id}
\mathscr{P} \exp \bigg( - \int_0^{\theta} d \tau \Gamma_{\theta} (\tau) \bigg) \mathscr{P} \exp \bigg(\int_0^{\theta} d \tau  \Gamma^{op}_{\theta} (-\tau) \bigg) = \mathds{1} \otimes \mathds{1} \ , 
\end{equation}
Using the theorem in Appendix \ref{useful_th} with $A(\tau) = i  \Gamma_{\theta} (\tau)$ and $B(\tau) = i \Gamma^{op}_{\theta} (- \tau)$, we have that
\begin{equation}
\Gamma_{\theta} (\theta) = \Gamma^{op}_{\theta} (- \theta) \ .
\end{equation}
A continuous deformation of the type 
\begin{equation}
\mathcal{A} = \mathds{1} + \epsilon \delta \mathcal{A} + \mathcal{O} (\epsilon^2 ) \ , 
\end{equation}
is still a solution of (\ref{phys_unit_A}) and (\ref{braid_unit_A}), provided that the deformation $\delta \mathcal{A}$ satisfies the following conditions
\begin{equation}
\delta \mathcal{A} = - \delta \mathcal{A}^{op} = - \delta \mathcal{A}^{\dagger} \ . 
\end{equation}

\subsubsection{Algebra invariance}
The algebra invariance condition is 
\begin{equation}
\label{algebra_inv}
\Delta^{op} ( \mathfrak{a} ) R ( \theta ) = R (\theta ) \Delta (\mathfrak{a} ) \ , \qquad\qquad
\forall \ \mathfrak{a} \in \mathfrak{K} \ , 
\end{equation}
where $\mathfrak{K}$ is a generic superalgebra. 
By taking the derivative with respect to $\theta$ on both sides of (\ref{algebra_inv}), we obtain
\begin{equation}
\label{deriv_algebra_inv}
\frac{d \Delta^{op} (\mathfrak{a})}{d \theta}  - R\frac{d\Delta(\mathfrak{a})}{d\theta}  R^{-1} = \Delta^{op}(\mathfrak{a}) \Gamma_{\theta} - \Gamma_{\theta} \Delta^{op} (\mathfrak{a}) \ . 
\end{equation}
In the case where the coproduct satisfy the following property 
\begin{equation}
\label{coprod_special}
\frac{d \Delta (\mathfrak{a})}{d \theta} = c_{\mathfrak{a}} \Delta (\mathfrak{a}) \ , \qquad\qquad \forall \ \mathfrak{a} \in \mathfrak{K} \ , 
\end{equation}
then (\ref{deriv_algebra_inv}) implies 
\begin{equation}
[ \Delta^{op} (\mathfrak{a}) , \Gamma_{\theta} ] = 0 \ . 
\end{equation}
This happens for instance for the relativistic massless scattering in $AdS_2$ and $AdS_3$ integrable string backgrounds. 

The algebra invariance (\ref{algebra_inv}) evaluated at $\theta = 0$ gives 
\begin{equation}
\label{alg_inv_A}
\Delta^{op}(\mathfrak{a})|_{\theta =0}\, \mathcal{A} = \mathcal{A} \Delta(\mathfrak{a})|_{\theta =0}  \ . 
\end{equation}
For $\mathcal{A} = \Pi$, (\ref{alg_inv_A}) becomes 
\begin{equation}
\label{alg_inv_Pi}
\Delta^{op}(\mathfrak{a}) = \Pi  \Delta(\mathfrak{a}) \Pi \ ,
\end{equation}
which is simply the definition of the opposite coproduct. However for $\mathcal{A} = \mathds{1}$, (\ref{alg_inv_A}) becomes 
\begin{equation}
\label{alg_inv_1}
\Delta^{op} (\mathfrak{a}) =  \Delta (\mathfrak{a}) \ , 
\end{equation}
which occours for instance in the $AdS_2$ and $AdS_3$ cases in the relativistic limit, when all braiding factors trivialise. For a generic integrable system, if the given algebra does not satisfy (\ref{alg_inv_1}), even in a particular regime, than the solution $\mathcal{A} = \mathds{1}$ must be discarded. 

\begin{table}[h]
\begin{center}
\begin{tabular}{c@{\hskip 1cm}c@{\hskip 1cm}c}
\hline
 & Equation for $\mathcal{A}$ & Equation for $\Gamma$ \\
\hline
& &  \\
Physical unitarity & $\mathcal{A} \mathcal{A}^{\dagger} = \mathds{1} \otimes \mathds{1}$ & $\Gamma (\theta) = - \Gamma^{\dagger} (\theta)$  \\
 & &  \\
Braiding unitarity &  $\mathcal{A} \mathcal{A}^{op} =\mathds{1} \otimes \mathds{1}$  &  \\
& &  \\
   & case $\mathcal{A} = \Pi $ & $\Gamma (\theta) = \Gamma (- \theta)$   \\
& &  \\
   &  case $\mathcal{A} = \mathds{1}$ & $\Gamma (\theta) = \Gamma^{op} (- \theta)$   \\
& &  \\
Algebra invariance, & $\Delta^{op}(\mathfrak{a})|_{\theta =0}\, \mathcal{A} = \mathcal{A} \Delta(\mathfrak{a})|_{\theta =0}$ & $[ \Delta^{op} (\mathfrak{a}) , \Gamma ] = 0$ \\
coproduct (\ref{coprod_special}) & &   \\
\hline
\end{tabular}
\end{center}
\caption{Summary of the conditions for $(\Gamma , \mathcal{A})$. }
\label{tab:conditions}
\end{table}

\section{Conclusions}

In this paper, we have studied whether the R-matrix governing massless non-relativistic scatterings in $AdS_2 \times S^2 \times T^6$ is invariant under the action of the $q$-deformed Poincar\'e superalgebra in $d=2$. 
Despite the behaviour of the boost action on the massless and massive R-matrices in the context of other integrable $AdS$ backgrounds \cite{Joakim, Andrea, Riccardo, Riccardo1}, we found that the R-matrix is neither invariant nor annihilated by the boost action. 
Nevertheless, we found that a linear combination of the boost coproduct and its opposite annihilates the R-matrix. This condition can be naturally expressed in terms of a new variable $\gamma$, in the sense that the R-matrix must only depend on the difference $\gamma_1 - \gamma_2$ associated with the two particles. 

In the relativistic limit, $\gamma$ tends to the rapidity $\theta$ modulo a logarithmically divergent term, which disappears when considering the difference $\gamma_1- \gamma_2$. We found that the non-relativistic massless R-matrices describing right-right and left-left scatterings written in the $\gamma$ variable are exactly reproduced by the relativistic massless R-matrices with $\theta$ replaced by $\gamma$. This feature is also confirmed in $AdS_3 \times S^3 \times T^4$, where we also have numerical evidence that for real momenta the non-relativistic dressing factor is correctly reproduced from the relativistic one via the minimal prescription above, with the numerical accuracy given in the text. 
We have checked that the non-relativistic crossing equation reduces, in terms of the variable $\gamma$, to the one satisfied by the Sine-Gordon dressing phase at a special value of the coupling, which is relevant in the relativistic case. It would be desirable to show in a purely analytic fashion, without relying on numerical computations, that the expression available in the literature \cite{Borsato:2016xns} for the non-relativistic dressing phase does indeed only depend on the difference of the $\gamma$ variables, and attains the precise Sine-Gordon form without the contribution from any CDD factors. Progress in this direction has recently been made in \cite{Fontanella:2019ury}, where the property of being of difference form in the $\gamma$ variables was derived purely analytically for the massless non-relativistic $AdS_3$ dressing phase constructed in \cite{Borsato:2016xns}, and the absence of CDD factors was motivated. This has shown that (\ref{difference_gamma}) is indeed satisfied exactly also by the dressing factor as well, and therefore no modifications to that equation occur as a consequence of introducing such factor.

Supported by this evidence, we conjectured the non-relativistic dressing factor of the R-matrix in $AdS_2 \times S^2 \times T^6$, which it is still unknown, and we checked that the conjectured solution satisfies the non-relativistic crossing equation.

Motivated by the geometric interpretation of the boost action on the R-matrix in \cite{Andrea}, we started the program of classifying all possible R-matrices associated with a given integrable model with a certain (super)algebra. 
The conjectured expression for the most general R-matrix is given in terms of a connection $\Gamma$ on a fibre bundle and a constant matrix $\mathcal{A}$, which is fixed by initial conditions.
The R-matrix must satisfy a set of equations (\emph{e.g.} physical unitarity, brading unitarity, crossing symmetry, Yang-Baxter equation, algebra invariance) and this in turns implies a set of equations for the pair $(\Gamma, \mathcal{A})$. In this paper, we made some progress towards the understanding of  physical unitarity, braiding unitarity and the algebra invariance in terms of $(\Gamma , \mathcal{A})$. The conditions obtained from this set of equations are listed in table \ref{tab:conditions} and we checked that they are satisfied for the set of known integrable systems in Appendix \ref{appx:integr_syst}. 

Studying the Yang-Baxter and crossing equations turns out to be more involved. The Yang-Baxter equation involves the scattering with a third particle, and potentially this might be implemented via extending the base space with an extra coordinate. A representation-dependent formulation of crossing symmetry involves to implement the notion of supertransposition on one particle subspace. This procedure however turns out to be quite involved. 
We plan to investigate further on this in the future. 

One of the implication of our conjecture in the context of AdS/CFT massless scatterings is that one can safely restrict to classify only R-matrices which depend only on one (real or complex) variable, i.e. the rapidity $\theta$. Once the classification is done in this simpler context, one immediately obtains also the classification of the massless non-relativistic R-matrices, simply via the substitution $\theta \rightarrow \gamma$. 
We leave for future work the task of exploring whether this minimal prescription can also be applied in the context of the Bethe ansatz.

\acknowledgments

\noindent We thank Zoltan Bajnok, Rouven Frassek, Antonio Pittelli, Alexander Schenkel, Matthias Staudacher, Jock McOrist, Andrea Prinsloo, Joakim Str\"omwall, Istvan Sz\'ecs\'enyi, Stijn van Tongeren and Martin Wolf for several interesting discussions. We thank very much Bogdan Stefa\'nski for illuminating discussions, for reading the manuscript, and for a very interesting discussion regarding an analytic proof of our conjecture.
A. F. is grateful to the Department of Physics and Astronomy of the University of Padova for hospitality at an early stage of this work. In particular, A. F. thanks Davide Cassani, Kurt Lechner, Pieralberto Marchetti, Marco Matone and Dima Sorokin for discussions and for providing a nice and stimulating environment in Padova. 
During part of this work, A. F. has been supported by the \emph{Angelo Della Riccia Foundation} Fellowship. 
A. T. thanks the STFC for support under the Consolidated Grant project nr. ST/L000490/1.

\vskip 0.5cm

\vskip 0.5cm
\noindent{\bf Data Management} \vskip 0.1cm

\noindent No additional research data beyond the data presented and cited in this work are
needed to validate the research  findings in this work.

\vskip 0.5cm

\appendix

\section{Quaternionic notation}
\label{appx:quaternionic}

Let us focus for definiteness on the R-matrix {\it Solution 3} of \cite{Andrea2} for the right-right scattering and for the choice $\alpha=1$. It turns out that there is a compact quaternionic-type notation one can introduce to rewrite this matrix, {\it i.e.} 
\begin{equation}
\label{Yang}
R_3 = \mathds{1} \otimes \sigma_3 + \mu \, \sigma_1 \otimes \check{\sigma}_1\ , \qquad \mu \equiv e^{- \frac{\theta}{2}}\ , \qquad \theta \equiv \theta_1 - \theta_2\ , 
\end{equation}
where $\theta_i$ is the rapidity of the $i$-th particle, with dispersion relation
\begin{equation}
E_i = p_i = e^{\theta_i} \ ,
\end{equation} 
and
\begin{equation}
\sigma_3 = \mathsf{E}_{11} - \mathsf{E}_{22}\ , \qquad \sigma_1 = \mathsf{E}_{12} + \mathsf{E}_{21}\ , \qquad \check{\sigma}_1 = \mathsf{E}_{12} - \mathsf{E}_{21}\ ,
\end{equation}
where 
\begin{equation}
|1\rangle = |\phi\rangle \ , \qquad |2\rangle = |\psi\rangle\ ,
\end{equation}
and $|\phi\rangle$ is a boson, $|\psi\rangle$ a fermion.
The matrices $\{ \sigma_3, \sigma_1, \check{\sigma}_3 \}$ satisfy the quaternion algebra,  
\begin{eqnarray}
&&\sigma_1^2 = \sigma_3^2 = \mathds{1} \ , \qquad \check{\sigma}_1^2 = -\mathds{1} \ , \qquad \sigma_1 \, \sigma_3 = - \sigma_3 \, \sigma_1 = - \check{\sigma}_1 \ , \nonumber \\
&& \check{\sigma}_1 \, \sigma_3 = - \sigma_3 \, \check{\sigma}_1 = - \sigma_1 \ , \qquad  \sigma_1 \, \check{\sigma}_1 = - \check{\sigma}_1 \, \sigma_1 = - \sigma_3 \ .  
\end{eqnarray}
Expression (\ref{Yang}) is reminiscent of spin-chain $R$-matrices of Yangian-type, and it does indeed possess a very special Yangian symmetry \cite{alea}.

\section{Integral representation of the dressing factor}
\label{appx:integral}
In this appendix we show that the dressing factor for the R-matrix \emph{Solution 3} of \cite{Andrea2} admits an integral representation.

The $R$-matrix we shall focus on initially, dubbed {\it Solution 3} in \cite{Andrea2}, is provided by the following formula, valid for arbitrary values of $\alpha$:
\begin{eqnarray}
R_3(\theta) =\begin{pmatrix}
\label{left-right}
1 & 0 & 0 & \mp\alpha^{-2} e^{-\frac{\theta}{2}}\\
0 & -1 & e^{-\frac{\theta}{2}} & 0 \\
0 & e^{-\frac{\theta}{2}} & 1 & 0 \\
\mp\alpha^2 e^{-\frac{\theta}{2}} & 0 & 0 & -1 
\end{pmatrix},
\end{eqnarray}
where the upper sign is for right-right, the lower sign for left-left. It satisfies {\it cross-unitarity} \cite{Borsato:2016xns} (cf. also \cite{Bernard:1991vq}), but does \emph{not} satisfy braiding-unitarity by itself. Rather, it satisfies a mixed braiding unitarity relation with the $R$-matrix dubbed {\it Solution 5} in \cite{Andrea2}:
\begin{eqnarray}
R_5 = \begin{pmatrix}
\label{right-left}
1 & 0 & 0 & \pm \alpha^{-2} e^{\frac{\theta}{2}}\\
0 & 1 & e^{\frac{\theta}{2}} & 0 \\
0 & e^{\frac{\theta}{2}} & -1 & 0 \\
\pm \alpha^2 e^{\frac{\theta}{2}} & 0 & 0 & -1 
\end{pmatrix},
\end{eqnarray}
for arbitrary values of $\alpha$, with the upper sign for right-right scattering, the lower sign for left-left. If we set $\alpha^2=1$ for simplicity, and focus on right-right for definiteness, the mixed braiding unitarity condition, already anticipated in \cite{Hoare:2014kma}, is given by:
\begin{equation}
R_5^{op}(-\theta)R_3(\theta) = (1+e^{-\theta}) \mathds{1}\otimes \mathds{1}.
\end{equation} 
This relation allows to determine the dressing factor to associate with the Solution 5, say, $\Omega_5(\theta)$, from the knowledge of the one associated with the Solution 3:
\begin{equation}
\label{gin}
\Omega_5(-\theta)\Omega_3(\theta) = \frac{1}{1+e^{-\theta}},
\end{equation}
which can be trivially solved for $\Omega_5(\theta)$.

Because of this relationship, in what follows, we will simply write $\Omega$ instead of $\Omega_3$.

The fact that we have a non-trivial massless right-right and left-left scattering (surviving the BMN limit, which is the regime we are effectively taking the string theory to) is a non-perturbative effect, in agreement with Zamolodchikov's picture of massless scattering, outlined for instance in \cite{Zamol2,Zamol21,Borsato:2016xns}. The mixed right-left scattering is instead trivial, which signals that what we are actually describing {\it via} this scattering problem is a critical theory possessing scale invarance.

Focusing on Solution 3 for right-right scattering, we notice that crossing symmetry is implemented as follows.We define the supertranspose of a matrix $M$ to be
\begin{equation}
M^{str}_{ij} = (-)^{ij+i} \, M_{ji},
\end{equation}
and the charge conjugation matrix to be 
\begin{equation}
C = \mbox{diag}(i,1),
\end{equation}
such that 
\begin{eqnarray}
-\mathfrak{Q}_{q} = C^{-1} \mathfrak{Q}_{-q}^{str} \, C, \qquad -\mathfrak{S}_{q} = C^{-1} \mathfrak{S}_{-q}^{str} \, C,\nonumber \\
\end{eqnarray}
where the crossing transformation is defined by
\begin{equation}
q \to - q, \qquad \theta \to i \pi + \theta.
\end{equation}

The $R$-matrix satisfies a combined {\it crossing - braiding unitarity} condition:
\begin{equation}
\label{cror}
R (\theta) \, \big[C^{-1}\otimes \mathds{1}\big] \, R^{str_1}(i \pi + \theta) \big[C \otimes \mathds{1}\big] =  \mathds{1} \otimes \mathds{1} \ .
\end{equation}
In order to fulfil (\ref{cror}), we will equip the solution with an appropriate {\it dressing factor}: 
\begin{equation}
R = \Omega(\theta) \, R_3. 
\end{equation}
Eq. (\ref{cror}) implies that this factor has to satisfy
\begin{equation}
\label{f}
\Omega (\theta) \Omega(\theta + i \pi) = \frac{e^{\frac{\theta}{2}}}{2 \cosh \frac{\theta}{2}} \equiv f(\theta).
\end{equation}
As a consistency check, it is easy to verify that the dressing factor of the Solution 5 S-matrix satisfies a similar condition:
\begin{equation}
\label{fin}
\Omega_5 (\theta) \Omega_5(\theta + i \pi) = \frac{e^{\frac{-\theta}{2}}}{2 \cosh \frac{\theta}{2}} = f(-\theta),
\end{equation}
which reduces to (\ref{f}) upon using (\ref{gin}). Furthermore, using these equations one can deduce that
\begin{equation}
\Omega_5(\theta) = \Omega_3( i \pi - \theta).
\end{equation}

In \cite{Andrea2}, a {\it minimal} dressing factor was proposed:
\begin{eqnarray}
\label{omega}
\Omega(\theta) = \frac{e^{\frac{\gamma^{EM}}{2}- \frac{i \pi }{8}+\frac{\theta}{4}}}{\sqrt{2 \pi}} \prod_{j=1}^\infty e^{-\frac{1}{2 j}} \, j \, \frac{\Gamma\Big(j-\frac{1}{2}+\frac{\theta}{2\pi i}\Big)\Gamma\Big(j-\frac{\theta}{2\pi i}\Big)}{\Gamma\Big(j+\frac{1}{2}-\frac{\theta}{2\pi i}\Big)\Gamma\Big(j+\frac{\theta}{2\pi i}\Big)} \ ,
\end{eqnarray}
with $\gamma^{EM}$ being the Euler-Mascheroni constant. The ``minimal" nature of $\Omega(\theta)$ is due to the fact that it displays neither poles nor zeroes in the physical strip $\mathfrak{Im} \theta \in (0,\pi)$. From the representation (\ref{omega}) one can see that the function $\Omega(\theta)$ is meromorphic in the entire complex plane, with poles occurring at the following points on the imaginary axis:
\begin{eqnarray}
&&\theta = - i \pi (1 + 2 n), \quad n = 0,1,2,... \quad \longrightarrow \, \, \, \, \, \mbox{pole of order} \, \, \, n+1 \ , \nonumber\\
&&\theta = 2 i \pi m, \quad m = 1,2,... \quad\qquad\quad\ \,\,  \longrightarrow \, \, \, \, \, \mbox{pole of order} \, \, \, m \ ,
\end{eqnarray}
and zeroes at the following points on the imaginary axis:
\begin{eqnarray}
&&\theta = i \pi (1 + 2 n), \quad n = 1,2,... \quad \longrightarrow \, \, \, \, \, \mbox{zero of order} \, \, \, n \ , \nonumber\\
&&\theta = - 2 i \pi m, \quad m = 1,2,... \quad\quad\, \longrightarrow \, \, \, \, \, \mbox{zero of order} \, \, \, m \ .
\end{eqnarray}
The factor $\Omega(\theta)$ is actually analytic, with neither zeroes nor poles, in the strip $\mathfrak{Im} \theta \in (-\pi, 2 \pi)$.

Using the property $z \, \Gamma(z) = \Gamma(z+1)$ one can show 
\begin{equation}
\label{g}
\Omega(\theta) \, \Omega(- \theta) = \frac{e^{- i \frac{\pi}{4}}}{2 \cosh \frac{\theta}{2}} \equiv g(\theta).
\end{equation}

By combining (\ref{f}) and (\ref{g}) one obtains
\begin{eqnarray}
\Omega(\theta) = \frac{f(\theta)}{\Omega(\theta+i\pi)} = f(\theta)\frac{\Omega(- \theta - i \pi)}{g(\theta + i \pi)} = \frac{f(\theta)}{g(\theta+i\pi)}\frac{f(-\theta - i\pi)}{f(-\theta)} \Omega(i \pi - \theta),
\end{eqnarray}
where at the last stage we have used
\begin{eqnarray}
\Omega(i \pi - \theta) = \frac{f(-\theta)}{\Omega(-\theta)} = \frac{f(-\theta)}{f(- \theta - i \pi)} \Omega(-\theta - i \pi).
\end{eqnarray}
Altogether, this implies
\begin{equation}
\frac{\Omega(\theta)}{\Omega(i \pi - \theta)} = \frac{f(\theta)}{g(\theta+i\pi)}\frac{f(-\theta - i\pi)}{f(-\theta)} = e^{\frac{\theta}{2} - i \frac{\pi}{4}}
\end{equation}

Finally, the condition of {\it physical unitarity} of the S-matrix
\begin{equation}
S(\theta)S^\dagger(\theta) = \mathds{1} \otimes \mathds{1} \qquad \theta \ \ \mbox{real}, 
\end{equation} 
reads for the Solution 3 as follows\footnote{Since the $S$-matrix can be related to the $R$-matrix via $S = P R P$, where $P$ is the matrix implementing the permutation on two-particle states, and $P$ is a unitary matrix - being real, symmetric and self-inverse - we see that unitarity of $R$ is equivalent to unitarity of $S$.}:
\begin{equation}
R_3(\theta)R_3^\dagger(\theta) = (1+e^{-\theta}) \ \mathds{1} \otimes \mathds{1} \qquad \theta \ \ \mbox{real}. 
\end{equation} 
This implies that the dressing factor ought to satisfy
\begin{equation}
\label{indeed}
\Omega(\theta)\Omega^*(\theta) = \frac{1}{1+e^{-\theta}} \qquad \theta \ \ \mbox{real}. 
\end{equation} 
Using the explicit solution (\ref{omega}) we have verified numerically that this is indeed the case: the expression (\ref{omega}) satisifes (\ref{indeed}), hence the S-matrix associated to Solution 3 is a unitary matrix for real momenta, {\it i.e.} it satisfies physical unitarity.  

Taking inspiration from \cite{Karo1}, we now manipulate the dressing factor $\Omega(\theta)$ into an alternative form, which is traditionally more suitable for instance in the calculation of form factors. To this purpose, we can use the so-called {\it Malmst\'en} representation of the Gamma function: 
integral representation
\begin{equation}
\label{Malm}
\Gamma (z) = \exp \int_0^\infty \frac{e^{-t}}{t}\bigg[(z-1) - \frac{1 - e^{-(z-1)t}}{1 - e^{-t}}\bigg] \ ,
\end{equation}
valid for $\mathfrak{Re} z >0$ - see also \cite{Weisz:1977ij, Bombardelli}. It is clear that we can use this representation only if the intersection of all the domains of the gamma functions appearing in (\ref{omega}) is non-empty, namely if, $\forall \, \, j = 1,...,\infty$,

\begin{eqnarray}
\notag
\mathfrak{Re} \Big(j-\frac{1}{2}+\frac{\theta}{2\pi i}\Big) > 0 \quad \mbox{and} \quad \mathfrak{Re} \Big(j-\frac{\theta}{2\pi i}\Big) > 0 \phantom{\ .}\\ 
\mbox{and} \quad \mathfrak{Re} \Big(j+\frac{1}{2}-\frac{\theta}{2\pi i}\Big) > 0 \quad \mbox{and} \quad \mathfrak{Re} \Big(j+\frac{\theta}{2\pi i}\Big) > 0 \ . \nonumber
\end{eqnarray} 
Since these real parts are all monotonically increasing with $j$, the intersection is dictated by the lowest value which is $j=1$, which produces
\begin{equation}
\label{dom}
\mathfrak{Im} \theta \in (- \pi, 2 \pi) \ .
\end{equation}
Notice that the physical strip $\mathfrak{Im} \theta \in (0,\pi)$ is entirely contained in the domain of validity (\ref{dom}). Working in the domain (\ref{dom}), bringing all contributions under one integral, after a series of simplifications, one gets
\begin{eqnarray}
\label{omega2}
\Omega(\theta) = \frac{e^{\frac{\gamma}{2}- \frac{\pi i }{8}+\frac{\theta}{4}}}{\sqrt{2 \pi}} \prod_{j=1}^\infty e^{-\frac{1}{2 j}} \, j \, \exp \int_0^\infty \frac{e^{-t}}{t}\bigg[-1 + \frac{\cosh \big(\frac{t}{4} - \frac{t \theta}{2 \pi i}\big)}{\cosh \frac{t}{4}} \, \, e^{t\big[\frac{3}{2} - j\big]}\bigg] \ . 
\end{eqnarray}
We cannot simply transform the infinite product of exponentials into the exponent of an infinite sum, because the resulting expression does not converge. Differentiating w.r.t. $\theta$ the $\log$ of (\ref{omega2}) allows to get rid of the diverging piece, at which point bringing the sum over $j$ inside the integral produces a simple geometric series. One then gets  
\begin{eqnarray}
\label{intego}
K(\theta) \equiv \frac{d \log \Omega(\theta)}{d\theta} = \frac{1}{4} + \frac{i}{\pi} \int_0^\infty dx \, \frac{\sinh \Big[x \big(1 - \frac{2 \theta}{i \pi }\big)\Big]}{\cosh x \, \sinh 2 x} \ , \qquad \mathfrak{Im} \theta \in (- \pi, 2 \pi) \ .
\end{eqnarray}

Now we have the task of reconstructing $\Omega(\theta)$ from its logarithmic derivative, namely
\begin{eqnarray}
\label{integdo}
\Omega(\theta) = B \exp \int_0^\theta d \beta \, K(\beta) = B \exp A(\theta)\ ,  
\end{eqnarray}
where
\begin{equation}
A(\theta) = \frac{\theta}{4} + \frac{1}{2} \int_0^\infty \frac{dx}{x} \frac{\cosh [x (1 - \frac{2 \theta}{i \pi})] - \cosh x}{\cosh x \, \sinh 2 x} \ ,
\end{equation}
and $B$ is a constant. At the last step, we have swapped the integral over $\beta$ with the one over $x$. We have also chosen to leave the constant term inside the integral to ensure convergence near $x=0$. 

The initial integration value is now fixed by reproducing any specific value of $\Omega$ obtained from the original expression (\ref{omega}), for instance
\begin{equation}
\Omega(0) = \frac{e^{\frac{\gamma}{2}- \frac{\pi i }{8}}}{\sqrt{2 \pi}} \prod_{j=1}^\infty e^{-\frac{1}{2 j}} \, j \, \frac{\Gamma\big(j-\frac{1}{2}\big)}{\Gamma\big(j+\frac{1}{2}\big)} = \frac{e^{- i \frac{\pi}{8}}}{\sqrt{2}} \ .
\end{equation}
This means that we have to set
\begin{equation}
B = \frac{e^{- i \frac{\pi}{8}}}{\sqrt{2}}\ .
\end{equation}

\section{$(\Gamma, \mathcal{A})$ for various relativistic models}
\label{appx:integr_syst}

\begin{itemize}
\item Sine-Gordon model (non supersymmetric, fully bosonic)

\begin{equation}
R = \begin{pmatrix}
1 & 0 & 0 & 0 \\
0 & \frac{\sinh (\frac{\pi\theta}{\xi})}{\sinh(\frac{\pi(i\pi -\theta)}{\xi})} & 
\frac{\sin ( \frac{\pi^2}{\xi} )}{\sin ( \frac{\pi(\pi + i \theta)}{\xi})} & 0 \\
0 & \frac{\sin ( \frac{\pi^2}{\xi} )}{\sin ( \frac{\pi(\pi + i \theta)}{\xi})} & 
\frac{\sinh (\frac{\pi\theta}{\xi})}{\sinh(\frac{\pi(i\pi - \theta)}{\xi})} & 0 \\
0 & 0 & 0 & 1 
\end{pmatrix} \ , 
\end{equation}
\begin{equation}
\mathcal{A} = \begin{pmatrix}
1 & 0 & 0 & 0 \\
0 & 0 & 1 & 0 \\
0 & 1 & 0 & 0 \\
0 & 0 & 0 & 1 
\end{pmatrix} = \Pi \ , 
\end{equation} 
\begin{eqnarray}
\notag
\Gamma_{\theta} &=& - \frac{i \pi \sin(\frac{2\pi^2}{\xi})}{\xi \cos(\frac{2\pi^2}{\xi}) - \xi \cosh(\frac{2\pi\theta}{\xi})}  \mathsf{E}_{11} \otimes \mathsf{E}_{22} 
- \frac{2i\pi \cosh(\frac{\pi \theta}{\xi}) \sin(\frac{\pi^2}{\xi})}{\xi \cos(\frac{2\pi^2}{\xi}) - \xi \cosh (\frac{2\pi \theta}{\xi})} \mathsf{E}_{21} \otimes \mathsf{E}_{12}  \\
\notag
&-& \frac{2i\pi \cosh(\frac{\pi \theta}{\xi}) \sin(\frac{\pi^2}{\xi})}{\xi \cos(\frac{2\pi^2}{\xi}) - \xi \cosh (\frac{2\pi \theta}{\xi})} \mathsf{E}_{12} \otimes \mathsf{E}_{21} 
- \frac{i \pi \sin(\frac{2\pi^2}{\xi})}{\xi \cos(\frac{2\pi^2}{\xi}) - \xi \cosh(\frac{2\pi\theta}{\xi})}  \mathsf{E}_{22} \otimes \mathsf{E}_{11} \ .\\ 
\end{eqnarray}
where $\mathsf{E}_{ij}$ are the matrices wit all $0$s, but $1$ in row $i$ and column $j$. $\Gamma_{\theta}$ is an \emph{even} function of $\theta$.

\item Non-relativistic Heisenberg XXX spin chain (non supersymmetric, fully bosonic)
\begin{equation}
R = \frac{u}{u - 1} \bigg( \mathds{1} + \frac{\Pi}{u} \bigg) \ , \qquad\qquad
u \equiv u_1 - u_2 \ , 
\end{equation}
\begin{equation}
\mathcal{A} = \Pi \ , \qquad\qquad
\Gamma_u =  \frac{1}{u^2 - 1} \bigg( \mathds{1} - \Pi \bigg) \ . 
\end{equation}
$\Gamma_u$ is an \emph{even} function of $u$.

\item Integrable superstring in $AdS_5 \times S^5$. Forcing the massless and subsequently the relativistic limit for the choice of right-right kinematics, i.e.
\begin{equation}
x_\pm = e^{\pm i \frac{p}{2}}, \qquad p \to \epsilon \, e^\theta, \qquad \epsilon \to 0, 
\end{equation}
on the massive R-matrix \cite{beis0} - written having eliminated all explicit coupling-constant dependence using the $x^\pm$ constraint - we obtain

{\footnotesize
\setcounter{MaxMatrixCols}{20}
\begin{equation}
\begin{pmatrix}
-1 & 0 & 0 & 0 &0 & 0 & 0 & 0 & 0 & 0 & 0 & 0 & 0 & 0 & 0 & 0 \\
0 & - A^2 & 0 & 0 & C & 0 & 0 & 0 & 0 & 0 & 0 & AB & 0 & 0 & - AB & 0 \\
0 & 0 & A & 0 & 0 & 0 & 0 & 0 & - B & 0 & 0 & 0 & 0 & 0 & 0 & 0 \\
0 & 0 & 0 & A & 0 & 0 & 0 & 0 & 0 & 0 & 0 & 0 & - B & 0 & 0 & 0 \\
0 & -C & 0 & 0 & - A^2 & 0 & 0 & 0 & 0 & 0 & 0 &- AB & 0 & 0 & AB & 0 \\
0 & 0 & 0 & 0 & 0 & -1 & 0 & 0 & 0 & 0 & 0 & 0 & 0 & 0 & 0 & 0 \\
0 & 0 & 0 & 0 & 0 & 0 & A & 0 & 0 & - B & 0 & 0 & 0 & 0 & 0 & 0 \\
0 & 0 & 0 & 0 & 0 & 0 & 0 & A & 0 & 0 & 0 &0 & 0 & - B & 0 & 0 \\
0 & 0 & -B & 0 & 0 & 0 & 0 & 0 & - A & 0 & 0 & 0 & 0 & 0 & 0 & 0 \\
0 & 0 & 0 & 0 & 0 & 0 & - B & 0 & 0 & - A & 0 & 0 & 0 & 0 & 0 & 0 & \\
0 & 0 & 0 & 0 & 0 & 0 & 0 & 0 & 0 & 0 & 1 & 0 & 0 & 0 & 0 & 0 \\
0 & - AB & 0 & 0 & AB & 0 & 0 & 0 & 0 & 0 & 0 & A^2 & 0 & 0 & B^2 & 0 \\
0 & 0 & 0 & - B & 0 & 0 & 0 & 0 & 0 & 0 & 0 & 0 & - A & 0 & 0 & 0 & \\
0 & 0 & 0 & 0 & 0 & 0 & 0 & - B & 0 & 0 & 0 & 0 & 0 & - A & 0 & 0 \\
0 & AB & 0 & 0 & - AB & 0 & 0 & 0 & 0 & 0 & 0 & B^2 & 0 & 0 & A^2 & 0 \\
0 & 0 & 0 & 0 & 0 & 0 & 0 & 0 & 0 & 0 & 0 & 0 & 0 & 0 & 0 & 1 \\ 
\end{pmatrix}
\end{equation}
}
where
\begin{eqnarray}
A = \tanh\frac{\theta}{2} \ , \qquad
B = \frac{1}{\cosh\frac{\theta}{2}} \ , \qquad
C = - \frac{2}{1 + \cosh \theta} \ .
\end{eqnarray}

\begin{equation}
\Gamma_{\theta} = - \frac{1}{2\cosh \frac{\theta}{2}}\begin{pmatrix}
0 & 0 & 0 & 0 &0 & 0 & 0 & 0 & 0 & 0 & 0 & 0 & 0 & 0 & 0 & 0 \\
0 & 0 & 0 & 0 & 0 & 0 & 0 & 0 & 0 & 0 & 0 & -1 & 0 & 0 & 1 & 0 \\
0 & 0 & 0 & 0 & 0 & 0 & 0 & 0 & - 1 & 0 & 0 & 0 & 0 & 0 & 0 & 0 \\
0 & 0 & 0 & 0 & 0 & 0 & 0 & 0 & 0 & 0 & 0 & 0 & - 1 & 0 & 0 & 0 \\
0 & 0 & 0 & 0 & 0 & 0 & 0 & 0 & 0 & 0 & 0 & 1 & 0 & 0 & -1 & 0 \\
0 & 0 & 0 & 0 & 0 & 0 & 0 & 0 & 0 & 0 & 0 & 0 & 0 & 0 & 0 & 0 \\
0 & 0 & 0 & 0 & 0 & 0 & 0 & 0 & 0 & - 1 & 0 & 0 & 0 & 0 & 0 & 0 \\
0 & 0 & 0 & 0 & 0 & 0 & 0 & 0 & 0 & 0 & 0 &0 & 0 & - 1 & 0 & 0 \\
0 & 0 & 1 & 0 & 0 & 0 & 0 & 0 & 0 & 0 & 0 & 0 & 0 & 0 & 0 & 0 \\
0 & 0 & 0 & 0 & 0 & 0 & 1 & 0 & 0 & 0 & 0 & 0 & 0 & 0 & 0 & 0  \\
0 & 0 & 0 & 0 & 0 & 0 & 0 & 0 & 0 & 0 & 0 & 0 & 0 & 0 & 0 & 0 \\
0 & 1 & 0 & 0 & -1 & 0 & 0 & 0 & 0 & 0 & 0 & 0 & 0 & 0 & 0 & 0 \\
0 & 0 & 0 & 1 & 0 & 0 & 0 & 0 & 0 & 0 & 0 & 0 & 0 & 0 & 0 & 0  \\
0 & 0 & 0 & 0 & 0 & 0 & 0 & 1 & 0 & 0 & 0 & 0 & 0 & 0 & 0 & 0 \\
0 & -1 & 0 & 0 & 1 & 0 & 0 & 0 & 0 & 0 & 0 & 0 & 0 & 0 & 0 & 0 \\
0 & 0 & 0 & 0 & 0 & 0 & 0 & 0 & 0 & 0 & 0 & 0 & 0 & 0 & 0 & 0 \\ 
\end{pmatrix}
\end{equation}
$\Gamma_{\theta}$ is \emph{even} in $\theta$. 

\end{itemize}

In all three cases described above, as well as in $AdS_3$ and $AdS_2$ if we disregard parts proportional to the identity, one has that $\Gamma_\theta$ anticommutes with $\Pi$.

\section{Geometric interpretation of $\Gamma_M^{op}$}
We recall that 
\begin{equation}
\Gamma_M^{op} = \Pi \, \Gamma_M \, \Pi \ . 
\end{equation}
We notice that in the massless non-relativistic cases in $AdS_3 \times S^3 \times T^4$ and $AdS_5 \times S^5$ superstring backgrounds, the following relations hold
\begin{equation}
\Gamma_1^{op} (p_2 , p_1) = \Gamma_2 (p_1 , p_2) \ , \qquad\qquad
\Gamma_2^{op} (p_2 , p_1 ) = \Gamma_1 (p_1 , p_2) \ ,   
\end{equation}
or equivalently, in a vector notation
\begin{equation}
\label{gamma_op_cond}
\begin{pmatrix}
\Gamma_1 (p_1, p_2) \\ 
\Gamma_2 (p_1, p_2)
\end{pmatrix}^{op} = 
\begin{pmatrix}
\Gamma_2 (p_2 , p_1) \\ 
\Gamma_1 (p_2 , p_1) 
\end{pmatrix} \ .
\end{equation}
Here we show that the RHS term of (\ref{gamma_op_cond}) can be generated as a consequence of a rotation of $\pi /2$ anticlockwise of the axes $(p_1, p_2)$ followed by an inversion of the new $p_2$ axis. 

For an anticlockwise rotation of angle $\theta = \pi / 2$ of the axes $(p_1, p_2)$, we have that 
\begin{equation}
\begin{pmatrix}
p_1'\\
p_2'
\end{pmatrix} = 
\begin{pmatrix}
0 & 1 \\
- 1 & 0 
\end{pmatrix}
\begin{pmatrix}
p_1 \\
p_2
\end{pmatrix} = 
\begin{pmatrix}
p_2 \\
- p_1
\end{pmatrix} \ . 
\end{equation}
The inversion of the $p_2'$ axis can be written as
\begin{equation}
\begin{pmatrix}
p_1'' \\
p_2''
\end{pmatrix} = 
\begin{pmatrix}
1 & 0 \\
0 & -1 
\end{pmatrix} 
\begin{pmatrix}
p_1' \\
p_2''
\end{pmatrix}=
\begin{pmatrix}
p_1' \\
- p_2'
\end{pmatrix} \ . 
\end{equation}
The corresponding transformation of the connection $\Gamma_M$ is 
\begin{equation}
\begin{pmatrix}
\Gamma_1'' \\
\Gamma_2''
\end{pmatrix}(p''_1, p''_2) = 
\begin{pmatrix}
1 & 0 \\
0 & -1
\end{pmatrix}
\begin{pmatrix}
0 & 1 \\
-1 & 0 
\end{pmatrix}
\begin{pmatrix}
\Gamma_1 \\
\Gamma_2
\end{pmatrix} (p_1 (p_1'', p_2'') , p_2 (p_1'', p_2'') ) \ ,
\end{equation}
where we have that
\begin{equation}
p_1 (p_1'', p_2'') = p_2'' \ , \qquad\qquad p_2 (p_1'', p_2'') = p_1 \ , 
\end{equation}
and
\begin{equation}
\begin{pmatrix}
1 & 0 \\
0 & -1
\end{pmatrix}
\begin{pmatrix}
0 & 1 \\
-1 & 0 
\end{pmatrix}
\begin{pmatrix}
\Gamma_1 \\
\Gamma_2
\end{pmatrix} = 
\begin{pmatrix}
\Gamma_2 \\
\Gamma_1
\end{pmatrix} \ . 
\end{equation}
This implies that
\begin{equation}
\begin{pmatrix}
\Gamma_1'' \\
\Gamma_2''
\end{pmatrix}(p''_1, p''_2) =
\begin{pmatrix}
\Gamma_2 \\
\Gamma_1
\end{pmatrix} (p_2'' , p_1'') \ , 
\end{equation} 
and therefore, by using (\ref{gamma_op_cond}), we have 
\begin{equation}
\begin{pmatrix}
\Gamma_1'' \\
\Gamma_2''
\end{pmatrix}(p''_1, p''_2) =
\begin{pmatrix}
\Gamma_1  \\ 
\Gamma_2 
\end{pmatrix}^{op} (p''_1, p''_2) \ . 
\end{equation}
This argument shows that whenever (\ref{gamma_op_cond}) is satisfied, the $^{op}$ operation can be interpreted as an anticlockwise rotation of $\pi/2$ of the frame of the fibre bundle base space, followed by an axis inversion.

\section{A useful theorem}
\label{useful_th}

\begin{theorem}
Suppose that the following equation holds for any $\theta$ 
\begin{equation}
\label{condition_th}
\mathscr{P} \exp \bigg( i \int_0^{\theta} d \tau A (\tau) \bigg)
\mathscr{P} \exp \bigg( i \int^0_{\theta} d \tau B(\tau) \bigg) = 
\mathds{1} \otimes \mathds{1} \ , 
\end{equation} 
for generic operators $A$ and $B$. Then 
\begin{equation}
A (\theta ) = B (\theta ) \ . 
\end{equation}
\end{theorem}
\begin{proof}
We shall first recall the following property
\begin{equation}
\bigg[ \mathscr{P} \exp \bigg( i \int_0^{\theta} d \tau A (\tau) \bigg) \bigg]^{-1} = \mathscr{P} \exp \bigg(-  i \int_0^{\theta} d \tau A (\tau) \bigg) \ . 
\end{equation}
Then equation (\ref{condition_th}) can be rewritten as
\begin{equation}
\label{condition_th_2}
\mathscr{P} \exp \bigg(-  i \int_0^{\theta} d \tau A (\tau) \bigg) = 
\mathscr{P} \exp \bigg(-  i \int_0^{\theta} d \tau B (\tau) \bigg) \ . 
\end{equation}
Differentiating both members of (\ref{condition_th_2}) with respect to $\theta$, we obtain 
\begin{equation}
- i A (\theta) \mathscr{P} \exp \bigg(-  i \int_0^{\theta} d \tau A (\tau) \bigg) = 
- i B (\theta) \mathscr{P} \exp \bigg(-  i \int_0^{\theta} d \tau B (\tau) \bigg) \ , 
\end{equation}
and by using (\ref{condition_th_2}), this in turns implies
\begin{equation}
A(\theta) = B (\theta) \ . 
\end{equation}
\end{proof}

\end{document}